\documentclass[conference]{IEEEtran}
\IEEEoverridecommandlockouts
\usepackage{cite}
\usepackage{amsmath,amssymb,amsfonts}
\usepackage{algorithmic}
\usepackage{graphicx}
\usepackage{textcomp}
\usepackage{xcolor}
\usepackage{lipsum}
\usepackage{multirow}
\usepackage{diagbox}
\usepackage{float}
\usepackage[ruled]{algorithm2e}

\def\BibTeX{{\rm B\kern-.05em{\sc i\kern-.025em b}\kern-.08em
    T\kern-.1667em\lower.7ex\hbox{E}\kern-.125emX}}
\begin{document}

    \title{Lightweight Strategy for XOR PUFs as Security Primitives for Resource-constrained IoT device\\

}

\author{\IEEEauthorblockN{1\textsuperscript{st} Gaoxiang Li}
\IEEEauthorblockA{\textit{Department of Computer Science} \\
\textit{Texas Tech University}\\
Lubbock, TX 79409, USA\\
email address or ORCID}
\and
\IEEEauthorblockN{2\textsuperscript{nd} Yu Zhuang}
\IEEEauthorblockA{\textit{Department of Computer Science} \\
\textit{Texas Tech University}\\
City, Country \\
email address or ORCID}
\and
\IEEEauthorblockN{3\textsuperscript{rd} Khalid T. Mursi}
\IEEEauthorblockA{\textit{ College of Computer Science and Engineering} \\
\textit{University of Jeddah}\\
City, Country \\
email address or ORCID}

}

\maketitle

\begin{abstract}

    Physical Unclonable Functions (PUFs) are promising security primitives for resource-constrained IoT devices. And the XOR Arbiter PUF (XOR-PUF) is one of the most studied PUFs, out of an effort to improve the resistance against machine learning attacks of probably the most lightweight delay-based PUFs \,--\, the Arbiter PUFs. However, recent attack studies reveal that even XOR-PUFs with large XOR sizes are still not safe against machine learning attacks. Increasing PUF stages or components and using different challenges for different components are two ways to improve the security of APUF-based PUFs, but more stages or components lead to more hardware cost and higher operation power, and different challenges for different components require the transmission of more bits during operations, which also leads to higher power consumption. In this paper, we present a strategy that combines the choice of XOR Arbiter PUF (XOR-PUF) architecture parameters with the way XOR-PUFs are used to achieve lightweights in hardware cost and energy consumption as well as security against machine learning attacks. Experimental evaluations show that with the proposed strategy, highly lightweight component-differentially challenged XOR-PUFs can withstand the most powerful machine learning attacks developed so far and maintain excellent intra-device and inter-device performance, rendering this strategy a potential blueprint for the fabrication and use of XOR-PUFs for resource-constrained IoT applications.

\end{abstract}


\section{Introduction}

    \subsection{Overview and Motivation}
  

      The Internet of Things (IoTs) has a wide and deep participation in business and everyday life, forming a variety of networks. Many of them place a premium on security to ensure the integrity of their communications. However, many network nodes, such as sensors and IoT devices, are resource constrained and cannot support traditional cryptographic protocols, which are not lightweight. Physically Unclonable Functions \cite{gassend2002controlled,gassend2002silicon,lee2004technique,suh2007physical} (PUFs) have the potential to provide a lightweight cryptography solution to the omnipresent resource-constrained IoT. Unlike traditional methods, PUFs rely on inherent variations within integrated circuits to provide unique responses. Due to this lightweight feature, PUFs are appealing for resource-constrained IoT device identification and authentication\cite{herder2014physical,becker2015pitfalls,miorandi2012internet,yu2016lockdown}.

      PUFs can be divided into two types: weak PUFs and strong PUFs\cite{herder2014physical}. Weak PUF has a limited challenge-response pairs (CRP) space, so it is suitable for cryptographic key generation. Strong PUF, on the other hand, has an exponentially huge CRP space, which is suitable for challenge-response authentication protocols. Arbiter PUF (APUF) and its variant, XOR Arbiter PUF (XOR-PUF), are the most common implementations of strong PUFs. 
      
      However, strong PUFs are not necessarily "strong" in terms of modeling attack resistance. Though physically unclonable, some PUFs are ``mathematically clonable'' in the sense that the responses of a PUF can be predicted by machine learning (ML) modeling attacks. Previous studies have demonstrated that APUF is extremely vulnerable to ML modeling attacks. And its most widely studied variant, XOR-PUF, which is proposed to improve ML modeling attack resistance, is still incapable of surviving most recent ML modeling attacks \cite{ruhrmair2013puf,ruhrmair2010modeling,ganji2015attackers,alkatheiri2017towards,alkatheiri2017experimental,aseeri2018machine,mursi2020fast,wisiolattackers}.  \par
  
      Increasing PUF stages or components and using different challenges for different components are two ways to improve the security of APUF-based PUFs. To begin, as the number of components or stages increases, the cost of hardware and operational power increases proportionately, rendering them unsuitable for resource-constrained IoT devices. Second, previous research \cite{yu2016lockdown,wisiol2019breaking, aseeri2018subspace,mursi2021experimental} established that XOR-PUF with different challenges for different components (CDC-XPUF) can provide significant resistance to machine learning modeling attacks within the same XOR-PUF architecture. However, the number of transmission bits required by CDC-XPUFs is high, increasing overall hardware overhead and operating power. As a result, the current CDC-XPUF design is still unsuitable for IoT devices with limited resources. Therefore, there is a rising concern that XOR-PUF designs may suffer as a result of PUF designers' being forced to provide security at the expense of dramatically increased overhead.
      
      To begin, as the number of components or stages increases, the cost of hardware and operational power increases proportionately, rendering them unsuitable for resource-constrained IoT devices. Second, previous research \cite{yu2016lockdown,wisiol2019breaking, aseeri2018subspace,mursi2021experimental} established that XOR-PUF with different challenges for different components (CDC-XPUF) can provide significant resistance to machine learning modeling attacks within the same XOR-PUF architecture.

  In this paper, we describe a new lightweight CDC-XPUF strategy that achieves low hardware cost and energy consumption while also providing security against ML modeling attacks. Inspired by the different effects on modeling attack resistance between components and stages, we take a different approach by combining a lightweight XOR-PUF architecture parameter strategy that reduces the number of stages while increasing the number of components with the option of using XOR-PUFs with component-differentially sub-challenges. Our experimental results show that,
  \begin{itemize}
      \item 
      by reducing the number of stages while increasing the number of components in conventional CDC-XPUF architecture, our lightweight CDC-XPUFs can maintain high modeling attack resistance while significantly reduce up to 90\% hardware cost.
  \end{itemize}
  
  \begin{itemize}
      \item 
      the required transmission bits of lightweight CDC-XPUFs can be reduced to the same level as traditional XOR-PUFs and still maintain exponentially many challenge-response pairs (CRPs).
  \end{itemize}
  
  \begin{itemize}
      \item 
      intra-device and inter-device performance evaluation on FPGA hardware implementations confirm the lightweight CDC-XPUFs could attain solid uniqueness, randomness and improved reliability performance.
  \end{itemize}
  

    \subsection{Background Information on PUFs}

    In order to clarify technical discussions in later sections, we will briefly describe the mechanism of the arbiter PUF, XOR-PUF, and CDC-XPUF in this subsection.\par
    
    \subsubsection{The arbiter PUFs}
    \begin{figure}[htbp]
        \centering
         \includegraphics[width=8cm]{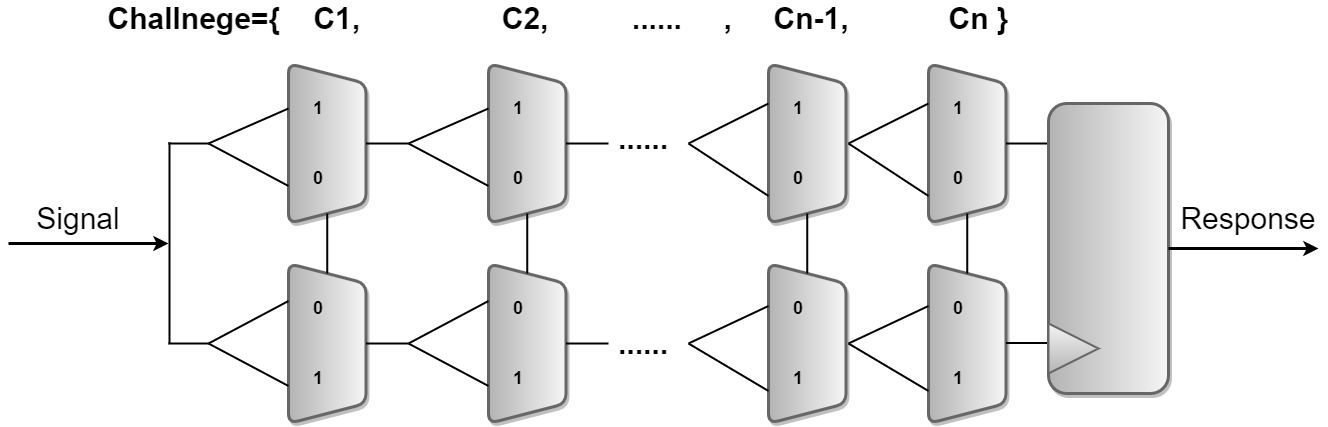}
      
        \caption{An aibiter PUF with n bits of challenge}
        \label{aibiter PUF}
    \end{figure}
    
    Fig.\ref{aibiter PUF} shows a simple case of an arbiter PUF. A $n$-bit arbiter PUF is made up of $n$ stages, each with two multiplexers (MUXs). When giving a rising signal, the signal enters the arbiter PUF from stage one and splits into two signals. The two signals are routed through gates at each stage, and the propagation paths are determined by the challenge bit to the multiplexers at each stage. Finally, two signals reach the D flip-flop, which acts as an arbiter to determine whether the signal on the top path or the signal on the lower path arrives first. If the top path signal arrives first, the D flip-flop returns $1$; otherwise, it returns $0$. \par

    \subsubsection{The XOR-PUFs and CDC-XPUFs}

    \begin{figure}[htbp]
        \centering
        \includegraphics[width=8cm]{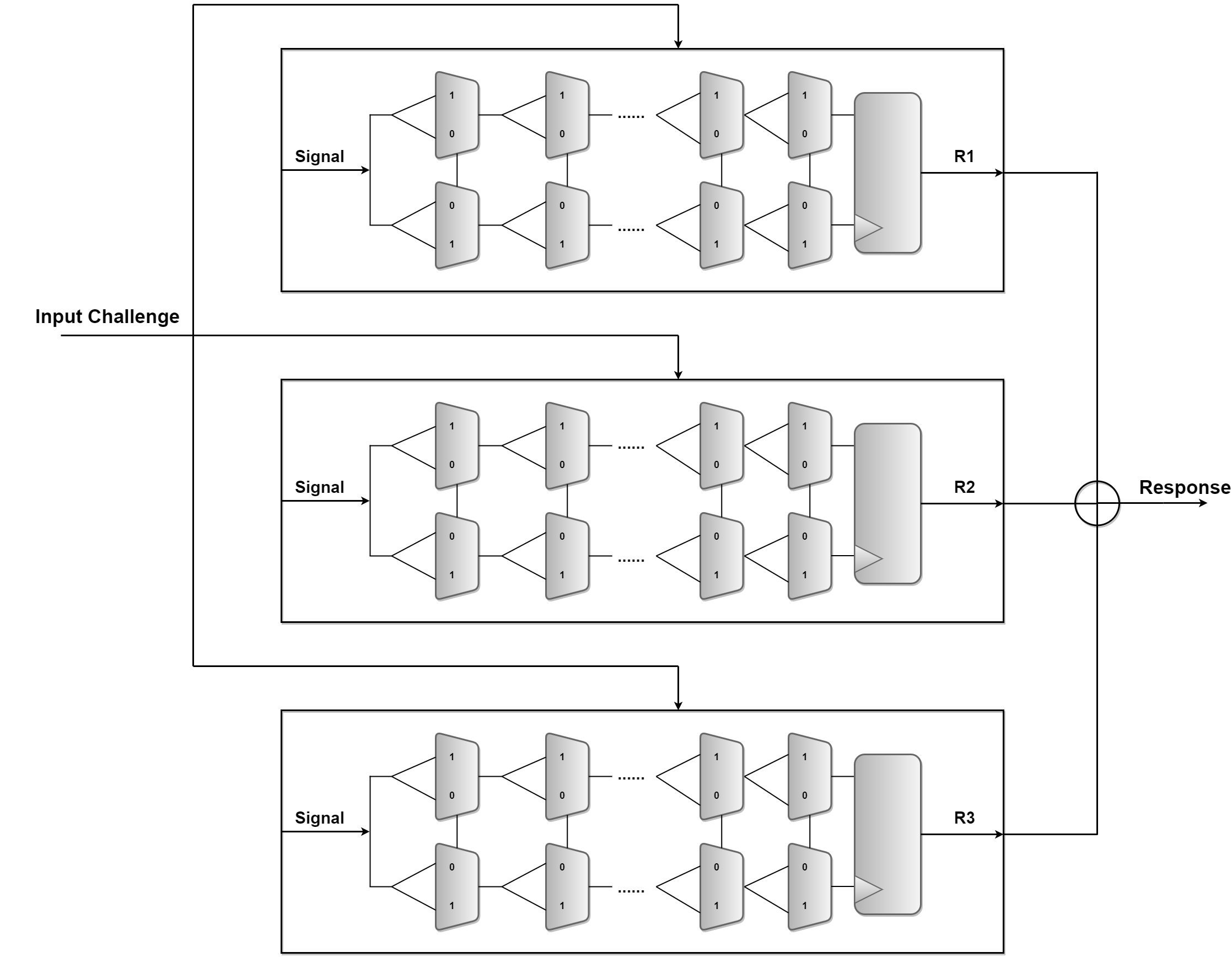}
        \caption{An XOR-PUF with 3 sub-stream and n bits of each stream}
        \label{XOR-PUF}
    \end{figure}
    
    Due to arbiter PUFs' weak resistance to ML modeling attacks, a new PUF was proposed in \cite{4261134} which increased a non-linear XOR gate to multiple arbiter PUFs to produce the final response. This type of PUF is known as the XOR arbiter PUF. Fig.\ref{XOR-PUF} illustrates a simple case of $n$-bit $3$-XOR-PUF. An \emph{n}-XOR-PUF is made up of \emph{n} component arbiter PUFs (also known as streams or sub-challenge) in which the responses of all \emph{n} component arbiter PUFs are XORed at XOR gate to produce one single bit response. It is worth noting that all component arbiter PUFs in an XOR-PUF are fed the same challenge bits.\par
    
    Studies in \cite{ruhrmair2010modeling,ruhrmair2013puf} show that XOR-PUFs could attain higher modeling attack resistance than arbiter PUFs. When equipped with lockdown scheme mutual authentication\cite{yu2016lockdown} to eliminate open-access interface, for XOR-PUFs with 64 stages and more than 9 component arbiter PUFs, all modeling attacks developed so far were not able to crack the XOR-PUF within the limited number of available CRPs (100 million). However, extending the number of streams and challenge stages will raise the cost and power consumption of a PUF, which is an important issue for resource-constrained IoT devices. Also, the expanding number of streams will lower the reliability of PUFs and increase the risk of reliability side-channel attacks\cite{becker2015gap}.\par
    
     Despite the fact that there are many alternative APUF variants and many new PUF designs proposed, such as Lightweight Secure PUFs \cite{majzoobi2008lightweight}, FF-PUFs \cite{gassend2004identification,lee2004technique, lim2004extracting}, and Interpose PUF \cite{nguyen2019interpose}, to the best of our knowledge, they are still vulnerable to ML modeling attacks \cite{tobisch2015scaling,santikellur2019deep,mursi2020MPUF_JCM,wisiol2020splitting,thapaliya2021machine}. In this paper, we will only focus on the most widely studied APUF variant \,--\, XOR-PUFs. 

    \begin{figure}[htbp]
        \centering
        \includegraphics[width=8cm]{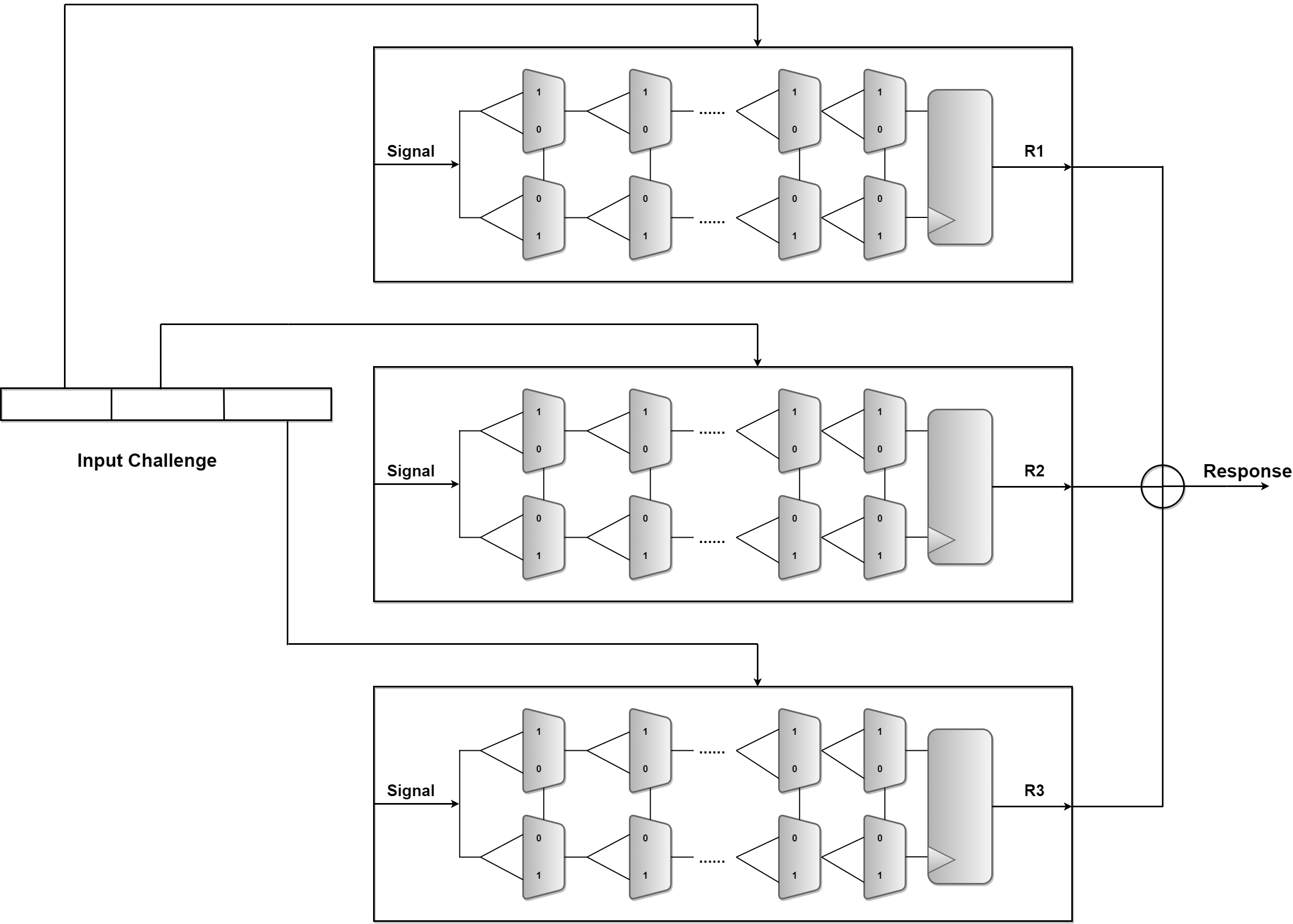}
        \caption{An XOR-PUF with 3 Component Differential Challenges and n bits of each challenges}
        \label{cdc PUF}
    \end{figure}

    For CDC-XPUFs, they share the same architecture as XOR-PUFs, which includes different multiple arbiter PUF components and XOR gates. The only difference between CDC-XPUF and XOR-PUF is that CDC-XPUF's each different component arbiter PUF receives different challenge inputs, while the XOR-PUF receives the same challenges for all its component arbiter PUFs. 

    Studies \cite{yu2016lockdown,wisiol2019breaking, aseeri2018subspace,mursi2021experimental} show that applying different challenges to different components of an XOR-PUF can decrease the vulnerability of the PUF against ML modeling attacks. Existing ML attack methods for 64-bit CDC-XPUFs with four components can attain a success rate lower than 90\% even if using more than one million CRPs. All the previous experimental results show that the 64-bit CDC-XPUF with four or more components is unbreakable or too expensive to break with existing attack methods. Therefore, these CDC-XPUFs can be considered as potentially good candidates in terms of security performance. The fact that CDC-XPUFs have high requirements for the number of transmission bits, on the other hand, increases the overall hardware overhead and operation power of CDC-XPUFs, and this issue has yet to be overcome. As a result, it is worthwhile to conduct a comprehensive study on CDC-XPUFs.
    
    As for another kind of malicious attack to PUFs, CDC-XPUFs are more vulnerable to reliability attacks\cite{becker2015gap,yu2016lockdown} than traditional XOR-PUFs. But this reliability attack can be prevented by applying a lockdown protocol to block the open interface. Although our lightweight CDC-XPUFs architecture will improve the reliability, it is not the purpose of this paper to investigate resistance to reliability attacks. In this paper, we will only focus on the machine learning modeling attack and assume that the lockdown authentication protocol is in place.

    \subsection{Organization of this Paper}
 
    The remaining of this paper is organized as follows. Section II gives a general overview of PUFs. The motivation for lightweight CDC-XPUFs design and its implementation are presented in section II. And the evaluation metric and tools for lightweight CDC-XPUFs will be presented in section III. In section IV, the experimental result of security evaluation based on simulator CRP and performance evaluation for lightweight CDC-XPUFs on simulator and FPGA implementations will be presented. Finally, concluding remarks are given in section V.\par

\section{Lightweight CDC-XPUFs design}

    
    As previously stated, CDC-XPUF is proposed to improve ML modeling attack resistance and maintain the architecture lightweight. However, the drawback that different challenges for different components in CDC-XPUFs require the transmission of more bits during operations, which also leads to higher power consumption. As a result, we are motivated to investigate a more lightweight CDC-XPUF to achieve low requirement of transmission bits. In this section, we will describe a new lightweight CDC-XPUF design strategy to reduce overall hardware cost while maintaining high ML attack resistance. 
    
  
\subsection{Factors impacting ML modeling attack resistance of APUF-based PUF}

    In general, there are two main factors influencing the ML modeling attack resistance of APUF-based PUF: the number of stages inside each arbiter PUF component and the number of component (the size of XOR-gate). To ensure security against ML attacks, both increasing the number of stages and increasing the number of components can improve the modeling attack resistance. However, the impact on the ML attack resistance of these two factors is not equivalent.
    
    For the first factor, inside the arbiter PUF component, the response $r$ of the additive delay model \cite{lim2004extracting}, which stipulates that the time it takes for each of the two signals to arrive at the arbiter are the summation of the delays incurred at all stages of the PUF. Based on the additive delay model, can be represented as
    \begin{equation}
    \label{equ:one}
    r = Sgn(v(n) + \sum_{i=1}^{n} w(i)\phi(i)) ,
    \end{equation}
    where $\phi$'s are transformed challenge \cite{lim2004extracting} given by
    \begin{equation}
    \label{equ:two}
    \phi(i) = (2c_i-1)(2c_{i+1}-1)\cdot\cdot\cdot\cdot\cdot(2c_n-1),
    \end{equation}
    with $c_i$ being the challenge bit at stage $i$, $v$ and $w$'s being parameters quantifying gate delays at different stages, and $Sgn(\cdot)$ the sign function. In (\ref{equ:one}), the term inside the $Sgn(\cdot)$ function is linear with respect to the transformed challenge $\phi$'s. The model represented by (\ref{equ:one}) is hence a \textbf{linear classification} problem with the separating hyperplane represented by equation
    \begin{equation}
    \nonumber
    w(1)\phi(1)+w(2)\phi(2)+\cdot\cdot\cdot\cdot\cdot+w(n)\phi(n)+v(n)=0,
    \end{equation}
    which results from setting to 0 the term inside the $Sgn(\cdot)$ function in (\ref{equ:one}). 
    
    For the second factor, inter the outputs from all arbiter PUF components, the response of the $k$-XOR arbiter PUF can be expressed as:
    \begin{equation}
    \textbf{\emph{r}} = \mathop{\bigoplus}\limits_{j = 1 \ldots k} r_{j},
    \label{eq_3}
    \end{equation} 
    where $r_{j}$ is the internal output of the $j^{th}$ component arbiter PUF. The XOR operation increases \textbf{non-linearity} of the relationship between the response $r$ and the transformed challenges $\phi$'s. Every additional arbiter PUF increases non-linearity as well as the dimension of the parameter space to be machine-learned by attackers \cite{ruhrmair2010modeling}, leading to higher resistance against machine learning attacks \cite{lim2005extracting}.
    
    As a rule, an non-linear classification problem is more difficult problem than the linear classification problem. As well, based on our experience and earlier reports about PUF modeling attack resistance, the ML modeling attack resistance of XOR-PUFs grows much greater as the number of components increases compared to the number of stages. Therefore, the number of components is a more important factor in ML modeling attack resistance of PUFs than the number of stages, while the number of stages affects attack resistance less effectively than the number of components.\par

    \subsection{Lightweight CDC-XPUF design}

    Guided by these two factors, for the purpose of maintaining high attack resistance and low hardware overhead, we could increase the number of components while decreasing the number of stages inside each component. And by this way, we are able to obtain a lightweight XOR-PUF design with high attack resistance in this manner. But as previously stated, XOR-PUFs are strong PUFs that should hold exponentially huge CRP spaces for identification and authentication applications. An $n$-stage $k$-XOR-PUF, which take the same challenge input for each component, can hold a $2^{n}$ available CRP space regardless of $k$. Thus, the number of challenge stages must be large for XOR-PUFs to maintain an exponentially huge CRP space. And this is the reason why the most common stage settings of traditional APUFs or XOR-PUFs are 64 or 128 bits. 
    
    However, because of the different challenges that are input into the different components of CDC-XPUFs, the available CRP space of CDC-XPUFs is much larger than that of XOR-PUFs. An $n$-stage CDC-$k$-XPUF can hold a $2^{n^K}$ available CRP space. To making a more detailed comparison, a 64-stage CDC-XPUF with 4 sub-challenges has a $2^{64^4}$ available CRP space, whereas a 64-stage XOR-PUF with 4 sub-challenges has a $2^{64}$ available CRP space. And both of these two CRP spaces are large enough to fulfill the authentication protocols. If the number of stages is reduced to 16, the available CRP space for a 16-stage CDC-XPUF with 4 sub-challenges is $2^{16^4}$ ($2^{64}$), which is still exponentially huge, whereas a 16-stage XOR-PUF with 4 sub-challenges only has a $2^{16}$ small available CRP space. Thus, considering the practical authentication and identification applications, XOR-PUFs are not suitable for this strategy to reduce hardware overhead due to the limited available CRP space caused by the stage-decreasing. This light-weighting method could only benefit CDC-XPUFs with enough CRP space.  \par

    To conclude this section, the number of components is a more important security factor than the number of stages in terms of PUF attack resistance. This observation motivates us to implement a new setting that increases the number of components while decreasing the number of stages to achieve high attack resistance while maintaining a low hardware cost. However, this method reduces the available CRP space of XOR-PUFs for future authentication protocol applications, and it is only applicable to CDC-XPUFs with a large CRP space. To address this issue, we are inspired to design the CDC-XPUF with shorter stages and more arbiter PUF components in order to keep the hardware overhead lightweight and resistant to modeling attacks.

\section{Evaluation of lightweight CDC-XPUFs}

    \subsection{Evaluation Tools for Modeling Attack Resistance }

    As is well known, the evaluation tool must exhibit strong attacking capabilities in order to make claims about the security of tested PUFs. We evaluate the resistance to machine learning attacks using the most powerful NN method \cite{mursi2020fast,wisiol2021neural} and the LR-based  method \cite{ruhrmair2013puf}. These methods have been widely used for PUF modeling attacking tasks and have been demonstrated to be one of the best.
    
    To demonstrate the practical effectiveness of our evaluation tools, we start with the evaluation of these evaluation tools by applying ML modeling attack on conventional CDC-XPUFs and XOR-PUFs with the NN method and the LR-based method. Tested PUFs contain 64-bit or 128-bit usual stage length. 
    
    Ten different simulated 64-bit or 128-bit CDC-XPUF and XOR-PUF instances for each number of components setting are generated using the Pypuf simulator library and CRP generator\cite{pypuf}. The generated PUF instances are all from the normal distribution, with a mean of 0 and a standard deviation of 1, and no noise value was added. The experiments employ a 90-10 training-testing split, with 1\% CRP from the training set used for validation. All the code is implemented in Python using the Tensorflow and Keras ML libraries \cite{abadi2016tensorflow,gulli2017deep}. The codes we developed for the experimental study, including codes implementing the CDC-XPUFs based on the Pypuf library, will be made available for reproducible study by peer researchers when the paper is published.\par
    
\begin{figure}[htbp]
        \centering
        \includegraphics[width=8cm]{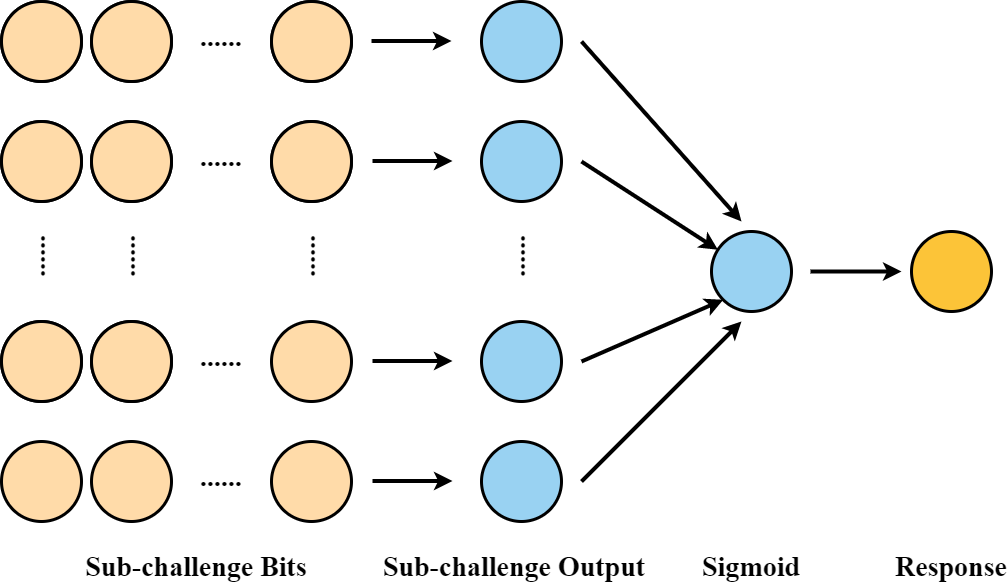}
        \caption{An overview of LR-based method for attacking CDC-XPUFs with component-differential-challenges}
        \label{overview}
    \end{figure}

\begin{figure}[htbp]
        \centering
        \includegraphics[width=8cm]{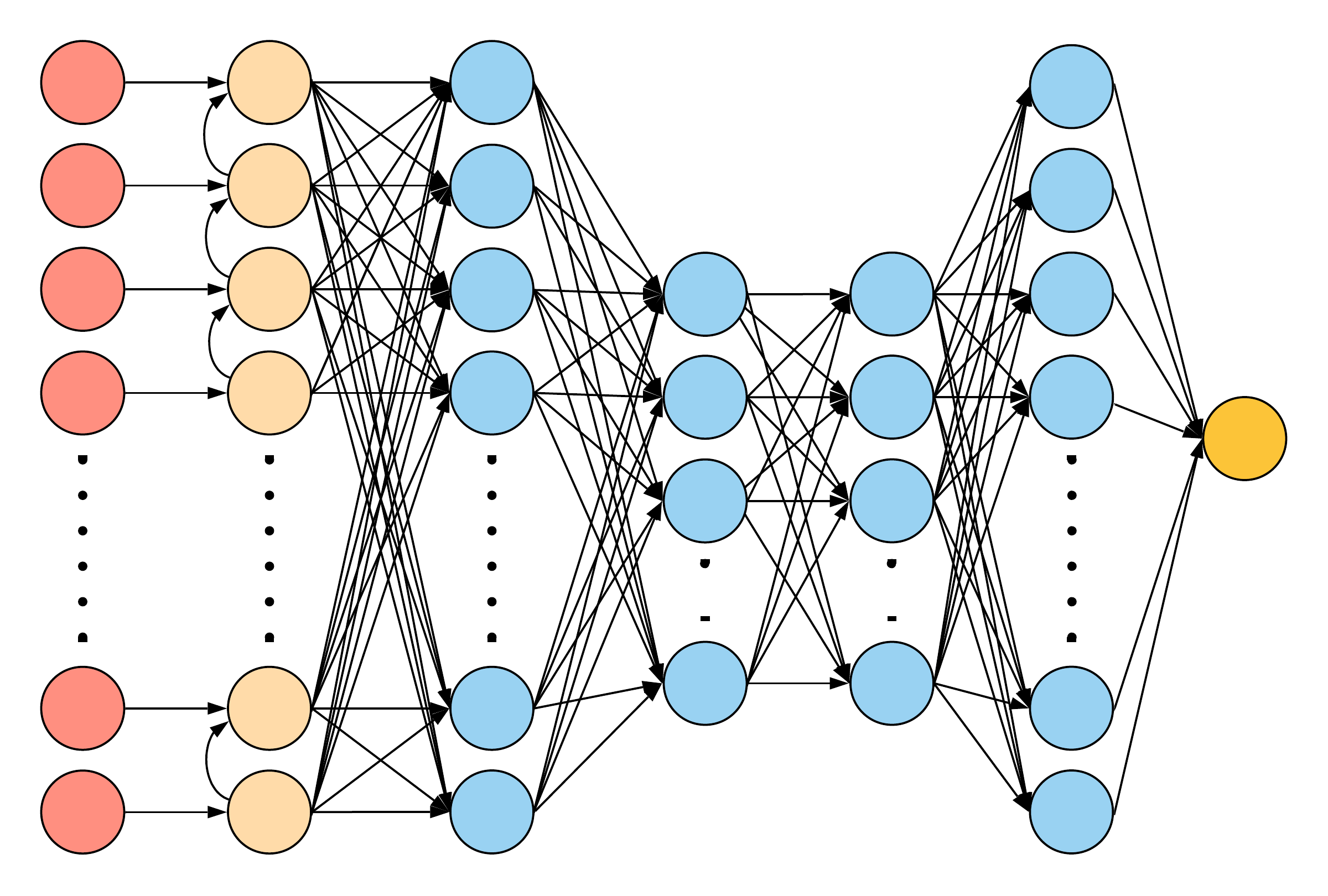}
        \caption{An overview of NN method for attacking CDC-XPUFs}
        \label{overview2}
    \end{figure}

    For the convenience of reading, the parameters of the NN attack method are listed in Table \ref{tab0} and the parameters for the LR-based method are listed in Table \ref{tab1}. An overview figure of the LR-based method is given in Fig.\ref{overview} and the overview of the NN method is given in Fig.\ref{overview2}. \par
   

\begin{table}[htbp]
\centering
\caption{Parameters of the NN attack method for $k$-XOR-PUFs and CDC-$k$-XPUFs  }
\linespread{1.3}\selectfont
    \setlength\tabcolsep{5pt}
\label{tab0}
\begin{tabular}{|c|c|c|c|c|}
\hline
{\textbf{Parameter}}&\textbf{Description}\\ \hline
\textbf{Optimizing Method}                             & ADAM \\ \hline
\textbf{Output Activation Function}                             &  Sigmoid    \\ \hline
\textbf{Learning Rate}                 &  Adaptive      \\ \hline
             & Layer 1 = Components $\times$ Stages  \\
   \textbf{No. Neurons in}             & Layer 2 = Components $\times$ Stages/ 2\\    
   \textbf{Each Layer}& Layer 3 = Components $\times$ Stages / 2 \\
   & Layer 4 = Components $\times$ Stages \\\hline
\textbf{Loss Function}                  &  Binary cross entropy       \\ \hline
\textbf{Batch Size}                  &  $10^{k-1}$       \\ \hline
\textbf{Kernel Initializer}      &Random Normal \\    \hline
\textbf{Early Stopping}      & True, when validation accuracy is 98\% \\    \hline
\end{tabular}
\end{table}

\begin{table}[htbp]
    \centering
    \caption{Parameters of the LR-based attack method for $k$-XOR-PUFs and CDC-$k$-XPUFs  }
    \linespread{1.3}\selectfont
    \setlength\tabcolsep{5pt}
    \label{tab1}
    \begin{tabular}{|c|c|c|c|c|}
    \hline
    {\textbf{Parameter}}&\textbf{Description}\\ \hline
    Optimizing Method                             & ADAM \\ \hline
    Output Activation Function.                             &  Sigmoid    \\ \hline
    Base Learning Rate                 &  0.01      \\ \hline
    Loss Function                  &  Binary cross entropy       \\ \hline
    Batch Size                  &  $10^{k-1}$       \\ \hline
    Patience                  &  5       \\ \hline
    
    \end{tabular}
\end{table}

\begin{table*}[htbp]
\caption{Modeling Attacking Performance Evaluation of the NN method }
\linespread{1.3}\selectfont
    \setlength\tabcolsep{5pt}
\label{tab:pufs_comaprision}

\centering
\begin{tabular}{|c|c|c|c|c|c|}
\hline
\textbf{Number of Stages}        & \textbf{PUF Type} & \textbf{Training Size} & \textbf{Average Accuracy} & \textbf{Training Time} & \textbf{Success Rate} \\ \hline
\multirow{10}{*}{\textbf{64 bits}} & \textbf{3 XOR-PUF}   & 5k                     & 99\%              & 10 sec                 & 100\%                 \\ \cline{2-6}
                                  & \textbf{4 XOR-PUF}   & 100k                    & 99\%              & 30 sec                 & 100\%                 \\ \cline{2-6} 
                                  & \textbf{5 XOR-PUF}   & 200k                    & 99\%              & 2 min                  & 100\%                 \\ \cline{2-6} 
                                  & \textbf{6 XOR-PUF}   & 1.2m                  & 99\%              & 3 min                  & 100\%                 \\ \cline{2-6} 
                                  & \textbf{7 XOR-PUF}   & 2m                   & 99\%              & 3 min                  & 100\%                 \\ \cline{2-6} 
                                  & \textbf{8 XOR-PUF}   & 10m                     & 99\%              & 5 min                  & 100\%                 \\ \cline{2-6} 
                                  & \textbf{9 XOR-PUF}   & 40m                     & 99\%              & 10 min                 & 90\%                  \\ \cline{2-6} 
                                 & \textbf{CDC-3-XPUF}   & 50k                     & 99\%              & 20 sec                 & 100\%                 \\ \cline{2-6}
                                  & \textbf{CDC-4-XPUF}          & 4m                   & 98\%              & 5 min                & 100\%                   \\ \cline{2-6} 
                                  & \textbf{CDC-5-XPUF}          & 80m                   & No Convergence              & 48 hours               & 0\%                   \\ \hline
\multirow{8}{*}{\textbf{128 bits}}   
                                  & \textbf{3 XOR-PUF}   & 25k                    & 98\%              & 30 sec                 & 100\%                 \\ \cline{2-6}
                                  & \textbf{4 XOR-PUF}   & 250k                    & 98\%              & 30 sec                 & 100\%                 \\ \cline{2-6} 
                                  & \textbf{5 XOR-PUF}   & 700k                    & 98\%              & 2 min                  & 100\%                 \\ \cline{2-6} 
                                  & \textbf{6 XOR-PUF}   & 10m                  & 98\%              & 3 min                  & 100\%                 \\ \cline{2-6} 
                                  & \textbf{7 XOR-PUF}   & 20m                   & 98\%              & 3 min                  & 100\%                 \\ \cline{2-6} 
                                  & \textbf{8 XOR-PUF}   & 80m                     & No Convergence              & 48 hours                  & 0\%                 \\ \cline{2-6} 
                                  & \textbf{CDC-3-XPUF}   & 400k                     & 96\%              & 5 min                  & 100\%                 \\ \cline{2-6}
                                 
                                   & \textbf{CDC-4-XPUF}          & 30m                   & 98\%              & 30 min                & 80\%                 \\ \hline
\end{tabular}
\end{table*}

\begin{table*}[htbp]
\caption{Modeling Attacking Performance Evaluation of the LR-based method }
\linespread{1.3}\selectfont
    \setlength\tabcolsep{5pt}
\label{tab:pufs_comaprision2}
\centering
\begin{tabular}{|c|c|c|c|c|c|}
\hline
\multicolumn{1}{|l|}{\textbf{Number of Stages}} &
  \multicolumn{1}{l|}{\textbf{PUF Type}} &
  \multicolumn{1}{l|}{\textbf{Training Size}} &
  \multicolumn{1}{l|}{\textbf{Average Accuracy}} &
  \multicolumn{1}{l|}{\textbf{Training Time}} &
  \multicolumn{1}{l|}{\textbf{Success Rate}} \\ \hline
\multirow{11}{*}{\textbf{64 bits}} & \textbf{3 XOR-PUF}                      & 4k   & 95\%           & 2 sec    & 100\% \\ \cline{2-6} 
                                   & \textbf{4 XOR-PUF}                      & 25k  & 97\%           & 10 sec   & 100\% \\ \cline{2-6} 
                                   & \textbf{5 XOR-PUF}                      & 400k & 97\%           & 20 sec   & 100\% \\ \cline{2-6} 
                                   & \textbf{6 XOR-PUF}                      & 2.5m & 98\%           & 2 min    & 90\%  \\ \cline{2-6} 
                                   & \textbf{7 XOR-PUF}                      & 10m  & 98\%           & 30 min   & 90\%  \\ \cline{2-6} 
                                   & \textbf{8 XOR-PUF}                      & 40m  & 96\%           & 1 hr     & 90\%  \\ \cline{2-6} 
                                   & \textbf{9 XOR-PUF}                      & 90m  & No Convergence & 48 hrs   & 0\%   \\ \cline{2-6} 
                                   & \multicolumn{1}{l|}{\textbf{CDC-3XPUF}} & 6k   & 96\%           & 10 sec   & 100\% \\ \cline{2-6} 
                                   & \multicolumn{1}{l|}{\textbf{CDC-4XPUF}} & 80k  & 97\%           & 1 min    & 100\% \\ \cline{2-6} 
                                   & \multicolumn{1}{l|}{\textbf{CDC-5XPUF}} & 4.5m & 96\%           & 30 min   & 90\%  \\ \cline{2-6} 
                                   & \multicolumn{1}{l|}{\textbf{CDC-6XPUF}} & 100m & 95\%           & 20 hours & 10\%  \\ \hline
\multirow{8}{*}{\textbf{128 bits}} & \textbf{3 XOR-PUF}                      & 10k  & 96\%           & 15 sec   & 100\% \\ \cline{2-6} 
                                   & \textbf{4 XOR-PUF}                      & 150k & 96\%           & 1 min    & 100\% \\ \cline{2-6} 
                                   & \textbf{5 XOR-PUF}                      & 2m   & 95\%           & 3 min    & 100\% \\ \cline{2-6} 
                                   & \textbf{6 XOR-PUF}                      & 35m  & 97\%           & 40 min   & 90\%  \\ \cline{2-6} 
                                   & \textbf{7 XOR-PUF}                      & 40m  & No Convergence & 24 hrs   & 0\%   \\ \cline{2-6} 
                                   & \multicolumn{1}{l|}{\textbf{CDC-3XPUF}} & 50k  & 96\%           & 10 sec   & 100\% \\ \cline{2-6} 
                                   & \multicolumn{1}{l|}{\textbf{CDC-4XPUF}} & 450k & 97\%           & 2 min    & 100\% \\ \cline{2-6} 
                                   & \multicolumn{1}{l|}{\textbf{CDC-5XPUF}} & 40m  & 97\%           & 2 hrs    & 50\%  \\ \hline
\end{tabular}
\end{table*}

    The evaluation results of the NN method and LR-based method on normal-stage-length XOR-PUFs and CDC-XPUFs are listed in Table \ref{tab:pufs_comaprision} and \ref{tab:pufs_comaprision2}. The results show that the NN attacking method is effective for breaking XOR-PUFs. Within given CRPs, all 64-bit XOR-PUFs with less than 9 sub-challenges can be cracked. It is even possible to crack 128-bit XOR-PUFs with fewer than 8 sub-challenges. In contrast, the LR-based method attain better performance than the NN method when attacking CDC-XPUFs. LR-based method can easily break CDC-XPUFs with 4 or 5 components and even could break 64-bit CDC-XPUFs with 6 components.\par
    
    Based on the result, the evaluation tools we used are powerful enough for examining the security vulnerability of our lightweight CDC-XPUFs against ML modeling attacks.

    \subsection{Performance Evaluation Metric for Lightweight CDC-XPUFs}

    When compared to conventional XOR-PUFs and CDC-XPUFs, the lightweight CDC-XPUF architecture significantly reduces the number of stages, raising the question of whether the new architecture could still achieve acceptable inter- and intra-device performance on the hardware implementation. This subsection will detail the PUF quantitative performance indicators on the hardware implementation, along with their definitions.

\subsubsection{Uniqueness of Responses of Different Chips when Inputting the Same Challenges}

    When the PUF was first introduced in 2002, one of its most notable features was its ability to produce different responses from different devices when presented with the same challenge. This property was later referred to as the chip fingerprint or uniqueness. In order to evaluate the CDC-XPUF uniqueness, we generated CRPs from different FPGAs for the same set of challenges. We used the uniqueness proposed by Hori et al. \cite{hori2010quantitative}.
    
    Hori's uniqueness (HU), introduced in \cite{hori2010quantitative}, calculates the uniqueness of responses of the same PUF design and challenges but from different devices. Hori's uniqueness is defined as follows:
    \begin{equation}
    \label{equ:HUn}
    HU_k =  \frac{4}{N_r \times N^2 }\sum_{i=1}^{N_r}\sum_{j,m=1, j\ne m}^{N}
    (b_j,_i \oplus b_m,_i),
    \end{equation}
    
    where $N$ is the number of chips, $N_r$ is the response length, and $b_{j,i}$ and $b_{k,i}$ are the $i$-th response bit from the $j$-th and $m$-th PUF instances respectively. 
    

    where $N$ is the number of chips, $N_r$ is the response length, and $b_{j,i}$ and $b_{k,i}$ are the $i$-th response bit from the $j$-th and $m$-th PUF instances respectively. 
    

\subsubsection{Measuring PUFs Reliability}

    PUF outputs are expected to be persistence. However, many conditions and circumstances may affect the design reliability, such as aging, heating, and the voltage level of the input signals. To measure how correct are PUF's outputs, we used the Steadiness metric introduce by Hori et al in \cite{hori2010quantitative}.\par

    An ideal PUF is expected to output the same response when is given the same single challenge on the same chip. Thus, the studying of the a PUF reliability in term of the design output is needed. The Steadiness ($S$), as defined in \ref{equ:steadiness}, measures if the a response bit ($b$) has changed or last the same among $N_a$ times of the same challenges. The steadiness is defined as follows:
    \begin{equation}
    \label{equ:steadiness}
    S = 1 + \frac{1}{N_c}\sum_{k=1}^{N_c}\log_{2}\max\{ \frac{\sum_{j=1}^
    {N_a} b_{k,j}}{N_a}, 
                         1 - \frac{\sum_{j=1}^{N_a} b_{k,j}}{N_a} \}
    \end{equation}
    where $N_c$ is denotes the total number of challenges and $k$ is a single challenge.

\subsubsection{Difference of Responses from the same PUF instances to Different Challenges}

    In terms of reaching the unpredictability of PUF's outputs, each design is expected to produce different responses when given different challenges. If PUF's outputs were biased to 0 or 1, it would be easy for an attacker to guess the responses with a higher prediction rate. Thus, evaluating PUF responses in terms of randomness and steadiness, introduced in \cite{hori2010quantitative}, is essential for guaranteeing the needed authentication and key complexity levels when using PUFs in security applications.
    
    Randomness is the study of PUF responses balanced in terms of 0's and 1's. The ideal PUF is expected to produce responses that have a balance of 0's and 1s by 50\% for each when inputting different challenges to the same chip. We can calculate the frequency of 1's in responses of a chip for different challenges as follows:
    \begin{equation}
    \label{equ:p_n}
    p = \frac{1}{N_r } \sum_{i=1}^{N_r} b_i 
    \end{equation}
    where $N_r$ is the total number of responses and $b$ is the response bit. Then, Randomness ($H$) can be calculated as follows:
    \begin{equation}
    \label{equ:H_n}
    H = -\log_{2}\max(p, 1 - p)
    \end{equation}


\section{ Experimental Setup}

    \subsection{Experiment Settings for Modeling attacks}

    To begin with, we evaluate the modeling attack resistance of the lightweight CDC-XPUF design with both the NN model and the LR-based model by inputting a newly generated training set of simulated CRP and then predicting responses to new challenges. To generate new datasets, we reduce the number of stages within the component from the most common 64-bit or 128-bit setting to new 8, 16, and 24 bits settings for each component, while increasing the number of arbiter PUF components to implement our lightweight CDC-XPUF design. Using the Pypuf simulator and CRP generator, we generate 20 different simulated CDC-XPUF instances for each number of components setting with the same experiment settings in section III.A. \par
    
    In each attack, the number of CRPs used in the attack starts small and gradually increases until having reached a size (the size listed in the column "Training Size" in Table), which results in a 90\% attack success rate for all 20 PUF instances or a failure with 100 million CRPs. And only attacks with a testing accuracy of greater than 90\% are considered successful. Given that the final outputs of XOR gates may become less reliable as their size increases, we evaluate the attack resistance performance at 0\%, 2\%, and 5\% noisiness levels to identify the performance of our lightweight CDC-XPUF design in real-world applications.\par
    
    \subsection{FPGA implementation}

    In our performance evaluation experiments, we programmed the different CDC-XPUFs on Xilinx Artix\textregistered-7 FPGA, which includes a configurable MicroBlaze CPU. The experiments were repeated three times on three chips to be able to capture and evaluate the responses' uniqueness and the reliability experiments were repeated 256 times to generate 10M CRP on each instance to evaluate the steadiness performance. VHSIC Hardware  Description Language (VHDL) was used to construct the CDC-XPUFs on Xilinx Vivado 15.4 HL design edition. The CDC-XPUFs placement in the hardware was done horizontally using Tool Command Language (TCL). Also, the Xilinx SDK was used to pass random challenges to the CDC-XPUFs and receive the corresponding response of each challenge.

    For generating the CRPs, AXI General Purpose Input/Output (GPIO) interfaces were used as follows: 1 GPIO for submitting the initial traveling signals, 1 for receiving the output response, and the leftovers for feeding the CDC-XPUF with the generated challenges. Our challenges were generated using the Pseudo-Random Number Generator (PRNG) as follows:
    \begin{equation}
    \label{equ:PRNG}
    C_{n+1} = (a \times C_n + g) \mod m,
    \end{equation}
    where $C$ is the sequence of the generated random number, $a$ is a multiplier, $g$ is a given constant, and $m$ is $2^{K}$ where K is the number of stages. To speed up the data transfer between PUFs and the computer, AXI Universal Asynchronous Receiver Transmitter (UART) was used with bud rate equals to $230,400$ bits/second. Finally, the Tera Term, which is a terminal emulator program, used for printing and saving the device output. 
    
    Furthermore, the voltage are set to 2.0 W, and the junction temperature reported by the Xilinx Artix\textregistered-7 FPGA is  $26.0^\circ C $ and the Thermal maigin is ${59.0^\circ C}$(12.3W).

\section{ Experimental Results and Discussion of Lightweight CDC-XPUF}

    \subsection{Results of Modeling Attack Resistance Evaluation}
    
    Firstly, the ML modeling attack experimental results for lightweight CDC-XPUFs with shorter stages are listed in Table~\ref{tab_part1} and Table~\ref{tab_part2}. It is important to note that only attacks with greater than 90\% testing accuracy can be considered successful, and the term "Accuracy" refers to the average accuracy of successful attacks. For the PUF instance with a 0\% success rate, the term "Accuracy" refers to the average accuracy of all attacks\par

\begin{table*}[htbp]
    \centering
    \caption{Results of attacks on lightweight CDC-XPUF design with LR-based and NN methods(Part 1)  }
    \linespread{1.3}\selectfont
    \setlength\tabcolsep{5pt}
    \label{tab_part1}
   \begin{tabular}{|c|c|c|c|c|c|c|}
\hline
\textbf{Components} & \textbf{Stages}    & \textbf{Noise Level} & \textbf{Security Evaluator} & \textbf{Training Size} & \textbf{Accuracy}       & \textbf{Success Rate} \\ \hline
\multirow{12}{*}{\textbf{6}}  & \multirow{4}{*}{\textbf{8}}  & \textbf{0\%}         & \textbf{NN}                & \textbf{1.8m}          & \textbf{98\%}           & \textbf{100\%}        \\ \cline{3-7} 
                              &                              & \textbf{0\%}         & \textbf{LR}                & \textbf{190k}          & \textbf{99\%}           & \textbf{90\%}         \\ \cline{3-7} 
                              &                              & \textbf{2\%}         & \textbf{LR}                & \textbf{200k}          & \textbf{96\%}           & \textbf{100\%}        \\ \cline{3-7} 
                              &                              & \textbf{5\%}         & \textbf{LR}                & \textbf{300k}          & \textbf{91\%}           & \textbf{90\%}         \\ \cline{2-7} 
                              & \multirow{4}{*}{\textbf{16}} & \textbf{0\%}         & \textbf{NN}                & \textbf{36m}           & \textbf{98\%}           & \textbf{50\%}         \\ \cline{3-7} 
                              &                              & \textbf{0\%}         & \textbf{LR}                & \textbf{1.5m}          & \textbf{97\%}           & \textbf{100\%}        \\ \cline{3-7} 
                              &                              & \textbf{2\%}         & \textbf{LR}                & \textbf{1.5m}          & \textbf{95\%}           & \textbf{100\%}        \\ \cline{3-7} 
                              &                              & \textbf{5\%}         & \textbf{LR}                & \textbf{1.8m}          & \textbf{90\%}           & \textbf{90\%}         \\ \cline{2-7} 
                              & \multirow{4}{*}{\textbf{24}} & \textbf{0\%}         & \textbf{NN}                & \textbf{80m}           & \textbf{No Convergence} & \textbf{0\%}          \\ \cline{3-7} 
                              &                              & \textbf{0\%}         & \textbf{LR}                & \textbf{1.8m}          & \textbf{96\%}           & \textbf{90\%}         \\ \cline{3-7} 
                              &                              & \textbf{2\%}         & \textbf{LR}                & \textbf{2.3m}          & \textbf{95\%}           & \textbf{100\%}        \\ \cline{3-7} 
                              &                              & \textbf{5\%}         & \textbf{LR}                & \textbf{2.7m}          & \textbf{90\%}           & \textbf{90\%}         \\ \hline
\multirow{12}{*}{\textbf{7}}  & \multirow{4}{*}{\textbf{8}}  & \textbf{0\%}         & \textbf{NN}                & \textbf{18m}           & \textbf{98\%}           & \textbf{90\%}         \\ \cline{3-7} 
                              &                              & \textbf{0\%}         & \textbf{LR}                & \textbf{1.8m}          & \textbf{99\%}           & \textbf{100\%}        \\ \cline{3-7} 
                              &                              & \textbf{2\%}         & \textbf{LR}                & \textbf{2.3m}          & \textbf{94\%}           & \textbf{100\%}        \\ \cline{3-7} 
                              &                              & \textbf{5\%}         & \textbf{LR}                & \textbf{3.6m}          & \textbf{90\%}           & \textbf{90\%}         \\ \cline{2-7} 
                              & \multirow{4}{*}{\textbf{16}} & \textbf{0\%}         & \textbf{NN}                & \textbf{100m}          & \textbf{No Convergence} & \textbf{0\%}          \\ \cline{3-7} 
                              &                              & \textbf{0\%}         & \textbf{LR}                & \textbf{2.7m}          & \textbf{97\%}           & \textbf{90\%}         \\ \cline{3-7} 
                              &                              & \textbf{2\%}         & \textbf{LR}                & \textbf{3.1m}          & \textbf{94\%}           & \textbf{90\%}         \\ \cline{3-7} 
                              &                              & \textbf{5\%}         & \textbf{LR}                & \textbf{100m}          & \textbf{90\%}           & \textbf{20\%}         \\ \cline{2-7} 
                              & \multirow{4}{*}{\textbf{24}} & \textbf{0\%}         & \textbf{NN}                & \textbf{100m}          & \textbf{No Convergence} & \textbf{0\%}          \\ \cline{3-7} 
                              &                              & \textbf{0\%}         & \textbf{LR}                & \textbf{35m}           & \textbf{96\%}           & \textbf{90\%}         \\ \cline{3-7} 
                              &                              & \textbf{2\%}         & \textbf{LR}                & \textbf{55m}           & \textbf{95\%}           & \textbf{100\%}        \\ \cline{3-7} 
                              &                              & \textbf{5\%}         & \textbf{LR}                & \textbf{100m}          & \textbf{90\%}           & \textbf{10\%}         \\ \hline
\end{tabular}
    \end{table*}
    
  \begin{table*}[htbp]
    \centering
    \caption{Results of attacks on lightweight CDC-XPUF design with LR-based and NN methods(Part 2)  }
    \linespread{1.3}\selectfont
    \setlength\tabcolsep{5pt}
    \label{tab_part2}
   \begin{tabular}{|c|c|c|c|c|c|c|}
\hline
\textbf{Components} & \textbf{Stages}    & \textbf{Noise Level} & \textbf{Security Evaluator} & \textbf{Training Size} & \textbf{Accuracy}       & \textbf{Success Rate} \\ \hline
\multirow{10}{*}{\textbf{8}}  & \multirow{4}{*}{\textbf{8}}  & \textbf{0\%}         & \textbf{NN}                & \textbf{100m}          & \textbf{96\%}           & \textbf{30\%}         \\ \cline{3-7} 
                              &                              & \textbf{0\%}         & \textbf{LR}                & \textbf{2.3m}          & \textbf{99\%}           & \textbf{90\%}         \\ \cline{3-7} 
                              &                              & \textbf{2\%}         & \textbf{LR}                & \textbf{2.7m}          & \textbf{94\%}           & \textbf{100\%}        \\ \cline{3-7} 
                              &                              & \textbf{5\%}         & \textbf{LR}                & \textbf{100m}          & \textbf{90\%}           & \textbf{20\%}         \\ \cline{2-7} 
                              & \multirow{4}{*}{\textbf{16}} & \textbf{0\%}         & \textbf{NN}                & \textbf{100m}          & \textbf{No Convergence} & \textbf{0\%}          \\ \cline{3-7} 
                              &                              & \textbf{0\%}         & \textbf{LR}                & \textbf{60m}           & \textbf{98\%}           & \textbf{90\%}         \\ \cline{3-7} 
                              &                              & \textbf{2\%}         & \textbf{LR}                & \textbf{100m}          & \textbf{95\%}           & \textbf{40\%}         \\ \cline{3-7} 
                              &                              & \textbf{5\%}         & \textbf{LR}                & \textbf{100m}          & \textbf{88\%}           & \textbf{0\%}          \\ \cline{2-7} 
                              & \multirow{2}{*}{\textbf{24}} & \textbf{0\%}         & \textbf{NN}                & \textbf{100m}          & \textbf{No Convergence} & \textbf{0\%}          \\ \cline{3-7} 
                              &                              & \textbf{0\%}         & \textbf{LR}                & \textbf{100m}          & \textbf{No Convergence} & \textbf{0\%}          \\ \hline
\multirow{5}{*}{\textbf{9}}   & \multirow{4}{*}{\textbf{8}}  & \textbf{0\%}         & \textbf{NN}                & \textbf{100m}          & \textbf{No Convergence} & \textbf{0\%}          \\ \cline{3-7} 
                              &                              & \textbf{0\%}         & \textbf{LR}                & \textbf{18m}           & \textbf{98\%}           & \textbf{90\%}         \\ \cline{3-7} 
                              &                              & \textbf{2\%}         & \textbf{LR}                & \textbf{23m}           & \textbf{94\%}           & \textbf{100\%}        \\ \cline{3-7} 
                              &                              & \textbf{5\%}         & \textbf{LR}                & \textbf{100m}          & \textbf{88\%}           & \textbf{0\%}          \\ \cline{2-7} 
                              & \textbf{16}                  & \textbf{0\%}         & \textbf{LR}                & \textbf{100m}          & \textbf{No Convergence} & \textbf{0\%}          \\ \hline
\multirow{2}{*}{\textbf{10}}   & \multirow{2}{*}{\textbf{8}}  & \textbf{0\%}         & \textbf{LR}                & \textbf{100m}          & \textbf{99\%} & \textbf{10\%}          \\ \cline{3-7} 
                              &                              & \textbf{2\%}         & \textbf{LR}                & \textbf{100m}           & \textbf{No Convergence}           & \textbf{0\%}         \\ \cline{3-7} 
                             
                              \hline
\end{tabular}
    \end{table*}

    From Table~\ref{tab_part1} and Table~\ref{tab_part2}, the results show that lightweight CDC-XPUFs with shorter stages and more components could still achieve high resistance to modeling attack. The required CRPs to crack the previously mentioned 64-bit CDC-4-XPUFs are 80K, which is even worse than the performance of 8-bit CDC-6-XPUFs. Furthermore, the attack resistance of 64-bit CDC-5-XPUFs is lower than that of 8-bit CDC-9-XPUFs or 16-bit CDC-8-XPUFs, which require 4.5M CRPs to break. Overall, these results indicate that we can maintain high modeling attack resistance of CDC-XPUFs by acquiring shorter stages and more components architecture.\par 
    
    On the other hand, lightweight CDC-XPUFs with shorter stages and more components continue to dominate the exponentially large CRP space. The available CRP spaces for the lightweight CDC-XPUF design are listed in Table~\ref{tab10}.\par
    
    \subsection{Results of Performance Evaluation}

    Figures \ref{fig:Randomness.png}, \ref{fig:Steadiness.png}, and \ref{fig:Uniqueness.png} show the performance evaluation results of some Lightweight CDC-XPUFs with exceptional high ML attack resistance based on the security results. Those PUFs are 8-CDC-XPUF 16-bit, 8-CDC-XPUF 24-bit, 9-CDC-XPUF 16-bit, 9-CDC-XPUF 24-bit, 10-CDC-XPUF 8-bit, and 10-XPUF 64-bit as a comparision. All of the chosen PUFs attain high ML attack resistance and most of them were unbreakable against our attacking model and using 100M CRPs. 

    \subsubsection{Randomness Results}

\begin{figure}[t!]
	\centering
	\includegraphics[width=3in]{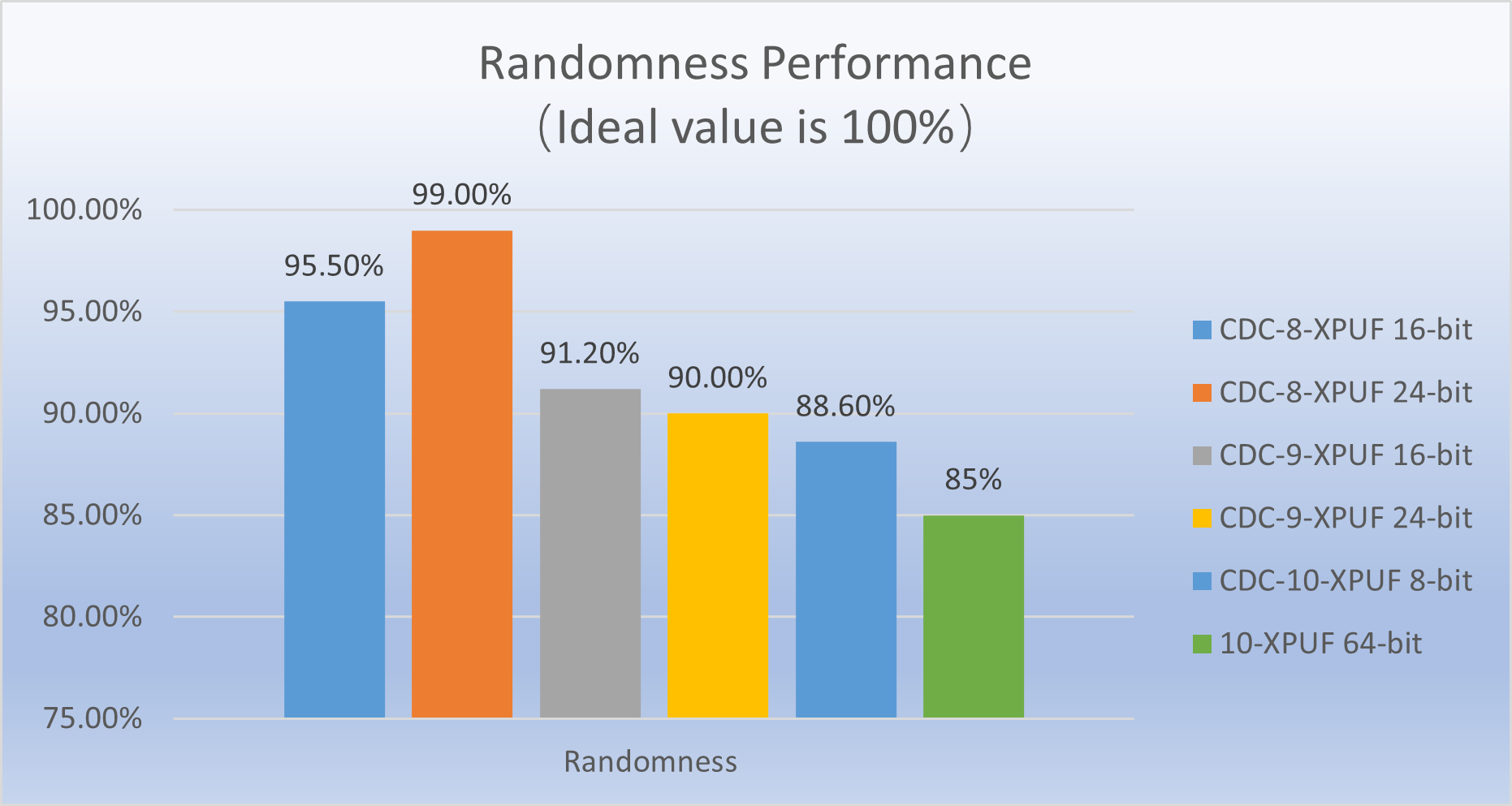}
	\caption{Calculating the difference of responses of the different challenges within the same chip. The ideal value of the Randomness is 100\%. }
	\label{fig:Randomness.png}
\end{figure} 

Figure \ref{fig:Randomness.png} shows the results of calculating the difference of the PUFs responses of different challenges and devices, equations \ref{equ:H_n}. The bar-graph shows the results of the randomness. Some CDC-XPUF shows strong randomness ranged from 95\% to 99\%, but some PUF shows a randomness around 90\%. This would be because the delays of the two selector chains are not idealy close, nevertheless equalizing the two delays is almost impracticable since the architecture of the FPGA is fixed. Nevertheless, in the face of an exponentially large available CRP space, the randomness performances of both CDC-XPUF and XOR-PUF are adequate for future applications.

    \subsubsection{Steadiness Results}

\begin{figure}[t!]
	\centering
	\includegraphics[width=3in]{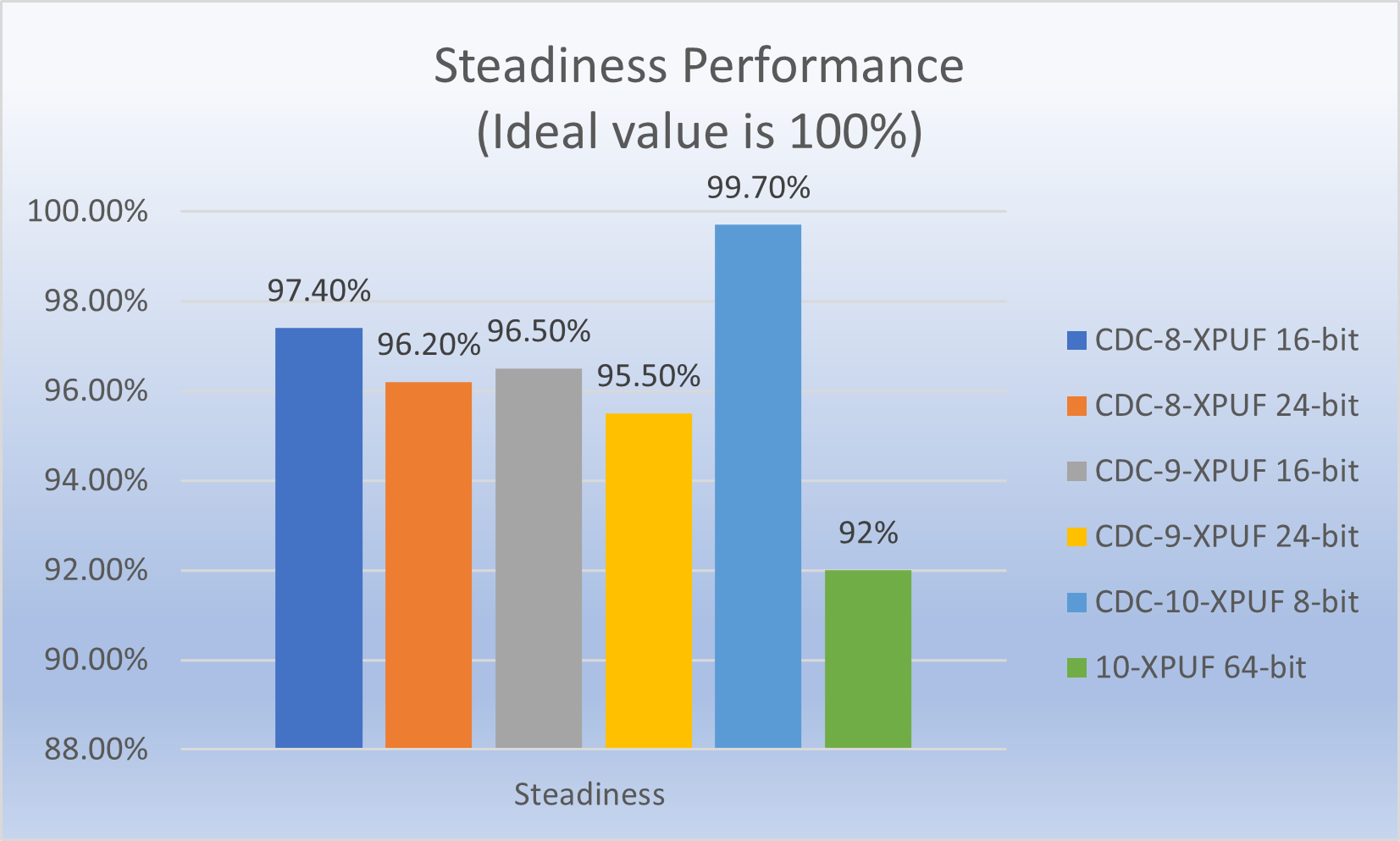}
	\caption{Calculating the reliability of CDC-XPUF. The ideal value of the Steadiness is 100\%.}
	\label{fig:Steadiness.png}
\end{figure} 
\begin{figure}[t!]
	\centering
	\includegraphics[width=3in]{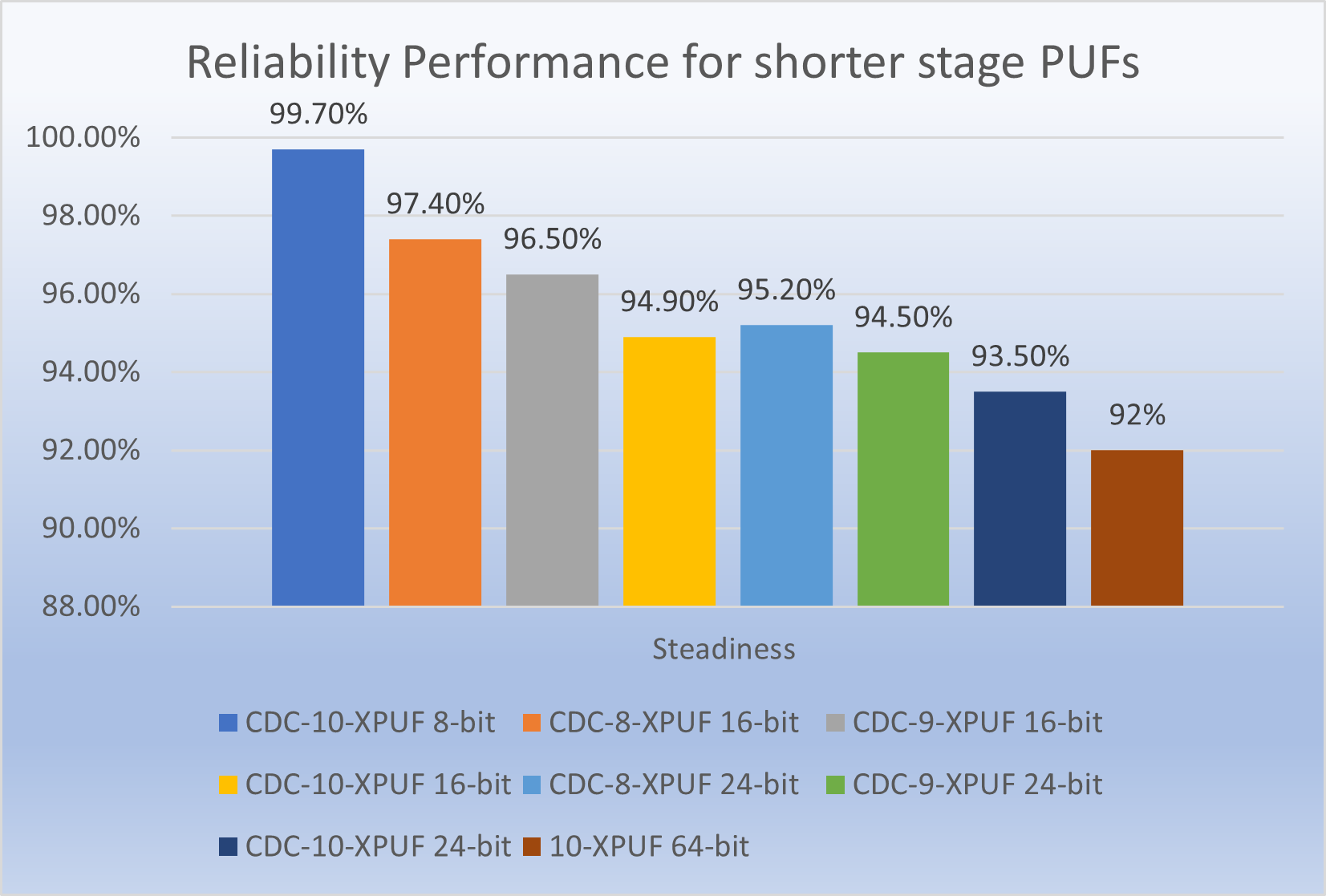}
	\caption{The reliability performance of PUFs with shorter stage. The ideal value is 100\%. The reliability decreases as the number of stages increases.}
	\label{fig:reliability.png}
\end{figure} 

Figure \ref{fig:Steadiness.png} illustrates the results of measuring the lightweight CDC-XPUF reliability, the difference of responses of the same challenge from the same chip. The steadiness bars are grouped on the figure. The lightweight CDC-XPUF shows a high steadiness of responses when repeating the responses with the same signal voltage and device temperature. 

Furthermore, based on Figure \ref{fig:reliability.png} we added two additional lightweight CDC-XPUF design int the comparison: 10-CDC-XPUF 16-bit and 10-CDC-XPUF 24-bit. We can observe the reliability of the tested PUF decrease as the number of stages increases. Therefore, the lightweight CDC-XPUF architecture with a shorter stage could help to improve the PUF reliability, which is a major expectation for an ideal PUF to output the same response when given the same single challenge on the same chip.

    \subsubsection{Uniqueness Results}

\begin{figure}[t!]
	\centering
	\includegraphics[width=3in]{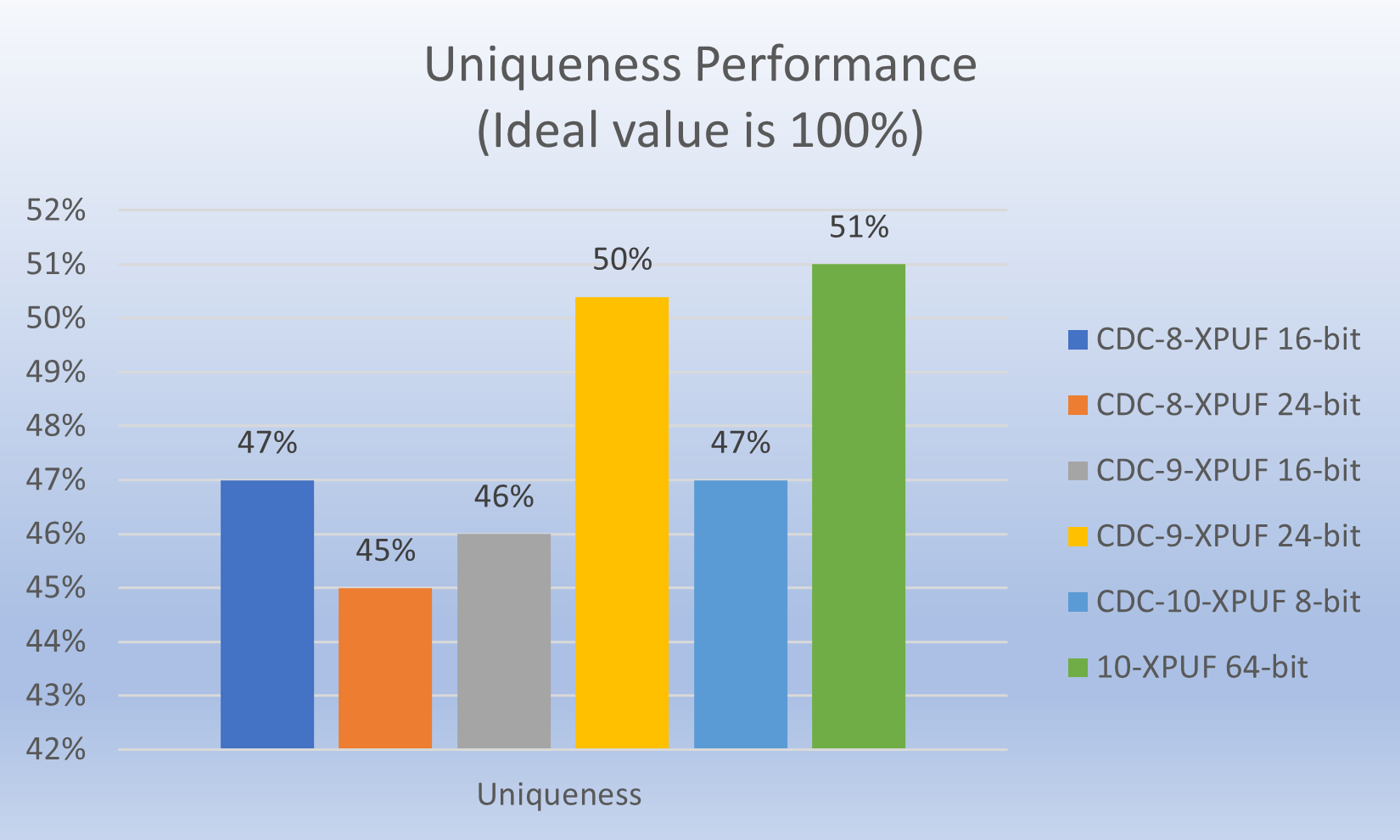}
	\caption{Calculating the uniqueness of the CDC-XPUF using Hori's equation. The ideal value of Hori's equation is 100\%. }
	\label{fig:Uniqueness.png}
\end{figure}

Figure \ref{fig:Uniqueness.png} demonstrates the uniqueness results which were calculated using Hori's equation \ref{equ:HUn}. The ideal value of the Hori's calculation is 100\%. Both CDC-XPUFs and XOR-PUFs show a uniqueness performance around 50\%. Nevertheless, when considering a huge range of possible challenges such as $2^{8\times10}$ in the CDC-10-XPUF 8-bit, 50\% of possible unique CRPs are relatively large and enough to consider the CDC-PUF in security applications.

    \subsection{ Hardware cost of lightweight CDC-XPUF design}
 
    Also, as previously stated, one of the drawbacks of CDC-XPUFs is that they require more overhead to transmit more bits. But the number of required bits during transmission is dramatically reduced by shortening the number of stages and increasing the number of arbiter components in CDC-XPUFs. Table~\ref{tab10} compares the required number of multiplexers (MUXs) and Arbiters hardware components, as well as the required transmission bits. We also include two interpose PUF instances in the comparison to evaluate the hardware cost.  The results, as shown in the table, indicate that the required number of transmission bits can be reduced to an acceptable level while maintaining high modeling attack resistance. In addition, the number of MUXs and Arbiters hardware components required for the lightweight CDC-XPUF design is reduced at the same time. On the other hand, the required hardware cost for the interpose PUF is much higher, and the modeling attack resistance is not as good as that of CDC-XPUFs. \par

    \begin{table*}[htbp]
        \caption{Overall hardware overhead in term of number of Multiplexers+Arbiters and required transmission bits that used for each PUF design(Each stage contains two MUXs)}
        \linespread{1.3}\selectfont
    \setlength\tabcolsep{5pt}
    \label{tab10}
    \centering
  \begin{tabular}{|c|c|c|c|c|c|c|}
\hline
\textbf{Components} & \textbf{Stages} & \textbf{PUF Type}                   & \textbf{Number of MUXs + Arbiters} & \textbf{Transmission bits} & \textbf{CRP Space} & \textbf{Required CRPs to crack} \\ \hline
\textbf{4}          & \textbf{64}     & \multirow{3}{*}{\textbf{XOR-PUF}}   & \textbf{$512+4$}             & \textbf{64}                & \textbf{$2^{64}$}  & \textbf{100k}                   \\ \cline{1-2} \cline{4-7} 
\textbf{5}          & \textbf{64}     &                                     & \textbf{$640+5$}             & \textbf{64}                & \textbf{$2^{64}$}  & \textbf{200k}                   \\ \cline{1-2} \cline{4-7} 
\textbf{9}          & \textbf{64}     &                                     & \textbf{$1152+9$}             & \textbf{64}                & \textbf{$2^{64}$}  & \textbf{40m}                    \\ \hline
\textbf{4}          & \textbf{64}     & \multirow{11}{*}{\textbf{CDC-XPUF}} & \textbf{$512+4$}             & \textbf{256}               & \textbf{$2^{256}$} & \textbf{80k}                    \\ \cline{1-2} \cline{4-7} 
\textbf{5}          & \textbf{64}     &                                     & \textbf{$640+5$}             & \textbf{320}               & \textbf{$2^{320}$} & \textbf{4.5m}                   \\ \cline{1-2} \cline{4-7} 
\textbf{6}          & \textbf{64}     &                                     & \textbf{$768+6$}             & \textbf{384}               & \textbf{$2^{384}$} & \textbf{100m}                   \\ \cline{1-2} \cline{4-7} 
\textbf{6}          & \textbf{8}      &                                     & \textbf{$96+6$}              & \textbf{48}                & \textbf{$2^{48}$}  & \textbf{190k}                   \\ \cline{1-2} \cline{4-7} 
\textbf{7}          & \textbf{8}      &                                     & \textbf{$112+7$}              & \textbf{56}                & \textbf{$2^{56}$}  & \textbf{1.8m}                   \\ \cline{1-2} \cline{4-7} 
\textbf{7}          & \textbf{24}     &                                     & \textbf{$336+7$}             & \textbf{168}               & \textbf{$2^{168}$} & \textbf{35m}                    \\ \cline{1-2} \cline{4-7} 
\textbf{8}          & \textbf{8}      &                                     & \textbf{$128+8$}              & \textbf{64}                & \textbf{$2^{64}$}  & \textbf{2.3m}                   \\ \cline{1-2} \cline{4-7} 
\textbf{8}          & \textbf{16}     &                                     & \textbf{$256+8$}             & \textbf{128}               & \textbf{$2^{128}$} & \textbf{60m}                    \\ \cline{1-2} \cline{4-7} 
\textbf{8}          & \textbf{24}     &                                     & \textbf{$384+8$}             & \textbf{192}               & \textbf{$2^{192}$} & \textbf{Over 100m}              \\ \cline{1-2} \cline{4-7} 
\textbf{9}          & \textbf{16}      &                                     & \textbf{$288+9$}              & \textbf{144}                & \textbf{$2^{144}$}  & \textbf{Over 100m}                   \\ \cline{1-2} \cline{4-7} 
\cline{4-7} 
\textbf{10}          & \textbf{8}      &                                     & \textbf{$160+10$}              & \textbf{80}                & \textbf{$2^{80}$}  & \textbf{Over 100m}                    \\ \hline

\textbf{8}          & \textbf{64}     & \textbf{(1,7)-IPUF}                 & \textbf{$519+8$} & \textbf{64}                & \textbf{$2^{64}$}  & \textbf{6m}                     \\ \hline
\textbf{14}         & \textbf{64}     & \textbf{(7,7)-IPUF}                 & \textbf{$903+8$} & \textbf{64}                & \textbf{$2^{64}$}  & \textbf{6m}                     \\ \hline
\end{tabular}
    \end{table*}
    
    These findings provide important evidence that the shortcomings of CDC-XPUFs, which cause more overhead to transmit more bits, can be overcome. As a result, not only can we maintain high modeling attack resistance by acquiring shorter stages and more component architecture, but we can also keep hardware costs and power consumption relatively low. To take it a step further, the lightweight CDC-XPUF design is a potentially good candidate for IoT device authentication because it combines high modeling attack resistance with low hardware cost.\par

\section{Conclusion}

     In this paper, we proposed a lightweight CDC-XPUF strategy for achieving high attack resistance while keeping operation and hardware costs, which further explores the potential of the current XOR-PUF architecture. This lightweight CDC-XPUF design can withstand modeling attacks while keeping the hardware overhead to a minimum. Unlike previous studies that focused on PUFs with 64 or 128 stages in each component, our research on CDC-XPUFs with short stages discovered that reducing the number of stages while increasing the number of arbiter PUF components could achieve good modeling attack resistance while keeping CDC-XPUFs lightweight in resource requirements. As a result, while CDC-XPUFs improve attack resistance at the expense of higher transmission overhead than XOR-PUFs, the transmission cost of this lightweight CDC-XPUF design can be kept to an acceptable level. Thus, reducing the number of stages while increasing the number of sub-challenges allows CDC-XPUFs to achieve high attack resistance while maintaining a low hardware cost. Furthermore, our study on performance evaluation also reveals that our lightweight CDC-XPUF design could attain solid uniqueness, randomness and improved reliability performance. This research will provide new insights into CDC-XPUFs and raises the possibility that XOR-PUF architectures can achieve high modeling attack resistance while also being low in hardware cost.
     \par

\section*{Acknowledgment}

\bibliographystyle{IEEEtran}
\bibliography{cite.bib}

\begin{thebibliography}{10}
\providecommand{\url}[1]{#1}
\csname url@samestyle\endcsname
\providecommand{\newblock}{\relax}
\providecommand{\bibinfo}[2]{#2}
\providecommand{\BIBentrySTDinterwordspacing}{\spaceskip=0pt\relax}
\providecommand{\BIBentryALTinterwordstretchfactor}{4}
\providecommand{\BIBentryALTinterwordspacing}{\spaceskip=\fontdimen2\font plus
\BIBentryALTinterwordstretchfactor\fontdimen3\font minus
  \fontdimen4\font\relax}
\providecommand{\BIBforeignlanguage}[2]{{%
\expandafter\ifx\csname l@#1\endcsname\relax
\typeout{** WARNING: IEEEtran.bst: No hyphenation pattern has been}%
\typeout{** loaded for the language `#1'. Using the pattern for}%
\typeout{** the default language instead.}%
\else
\language=\csname l@#1\endcsname
\fi
#2}}
\providecommand{\BIBdecl}{\relax}
\BIBdecl

\bibitem{gassend2002controlled}
B.~Gassend, D.~Clarke, M.~Van~Dijk, and S.~Devadas, ``Controlled physical
  random functions,'' in \emph{18th Annual Computer Security Applications
  Conference, 2002. Proceedings.}\hskip 1em plus 0.5em minus 0.4em\relax IEEE,
  2002, pp. 149--160.

\bibitem{gassend2002silicon}
B.~Gassend, D.~Clarke, M.~V. Dijk, and S.~Devadas, ``Silicon physical random
  functions,'' in \emph{Proceedings of the 9th ACM conference on Computer and
  communications security}, 2002, pp. 148--160.

\bibitem{lee2004technique}
J.~W. Lee, D.~Lim, B.~Gassend, G.~E. Suh, M.~Van~Dijk, and S.~Devadas, ``A
  technique to build a secret key in integrated circuits for identification and
  authentication applications,'' in \emph{2004 Symposium on VLSI Circuits.
  Digest of Technical Papers (IEEE Cat. No. 04CH37525)}.\hskip 1em plus 0.5em
  minus 0.4em\relax IEEE, 2004, pp. 176--179.

\bibitem{suh2007physical}
G.~E. Suh and S.~Devadas, ``Physical unclonable functions for device
  authentication and secret key generation,'' in \emph{2007 44th ACM/IEEE
  Design Automation Conference}.\hskip 1em plus 0.5em minus 0.4em\relax IEEE,
  2007, pp. 9--14.

\bibitem{herder2014physical}
C.~Herder, M.-D. Yu, F.~Koushanfar, and S.~Devadas, ``Physical unclonable
  functions and applications: A tutorial,'' \emph{Proceedings of the IEEE},
  vol. 102, no.~8, pp. 1126--1141, 2014.

\bibitem{becker2015pitfalls}
G.~T. Becker, ``On the pitfalls of using arbiter-pufs as building blocks,''
  \emph{IEEE Transactions on Computer-Aided Design of Integrated Circuits and
  Systems}, vol.~34, no.~8, pp. 1295--1307, 2015.

\bibitem{miorandi2012internet}
D.~Miorandi, S.~Sicari, F.~De~Pellegrini, and I.~Chlamtac, ``Internet of
  things: Vision, applications and research challenges,'' \emph{Ad hoc
  networks}, vol.~10, no.~7, pp. 1497--1516, 2012.

\bibitem{yu2016lockdown}
M.-D. Yu, M.~Hiller, J.~Delvaux, R.~Sowell, S.~Devadas, and I.~Verbauwhede, ``A
  lockdown technique to prevent machine learning on pufs for lightweight
  authentication,'' \emph{IEEE Transactions on Multi-Scale Computing Systems},
  vol.~2, no.~3, pp. 146--159, 2016.

\bibitem{ruhrmair2013puf}
U.~R{\"u}hrmair, J.~S{\"o}lter, F.~Sehnke, X.~Xu, A.~Mahmoud, V.~Stoyanova,
  G.~Dror, J.~Schmidhuber, W.~Burleson, and S.~Devadas, ``Puf modeling attacks
  on simulated and silicon data,'' \emph{IEEE transactions on information
  forensics and security}, vol.~8, no.~11, pp. 1876--1891, 2013.

\bibitem{ruhrmair2010modeling}
U.~R{\"u}hrmair, F.~Sehnke, J.~S{\"o}lter, G.~Dror, S.~Devadas, and
  J.~Schmidhuber, ``Modeling attacks on physical unclonable functions,'' in
  \emph{Proceedings of the 17th ACM conference on Computer and communications
  security}, 2010, pp. 237--249.

\bibitem{ganji2015attackers}
F.~Ganji, S.~Tajik, and J.-P. Seifert, ``Why attackers win: on the learnability
  of xor arbiter pufs,'' in \emph{International Conference on Trust and
  Trustworthy Computing}.\hskip 1em plus 0.5em minus 0.4em\relax Springer,
  2015, pp. 22--39.

\bibitem{alkatheiri2017towards}
M.~S. Alkatheiri and Y.~Zhuang, ``Towards fast and accurate machine learning
  attacks of feed-forward arbiter pufs,'' in \emph{2017 IEEE Conference on
  Dependable and Secure Computing}.\hskip 1em plus 0.5em minus 0.4em\relax
  IEEE, 2017, pp. 181--187.

\bibitem{alkatheiri2017experimental}
M.~S. Alkatheiri, Y.~Zhuang, M.~Korobkov, and A.~R. Sangi, ``An experimental
  study of the state-of-the-art pufs implemented on fpgas,'' in \emph{2017 IEEE
  Conference on Dependable and Secure Computing}.\hskip 1em plus 0.5em minus
  0.4em\relax IEEE, 2017, pp. 174--180.

\bibitem{aseeri2018machine}
A.~O. Aseeri, Y.~Zhuang, and M.~S. Alkatheiri, ``A machine learning-based
  security vulnerability study on xor pufs for resource-constraint internet of
  things,'' in \emph{2018 IEEE International Congress on Internet of Things
  (ICIOT)}.\hskip 1em plus 0.5em minus 0.4em\relax IEEE, 2018, pp. 49--56.

\bibitem{mursi2020fast}
K.~T. Mursi, B.~Thapaliya, Y.~Zhuang, A.~O. Aseeri, and M.~S. Alkatheiri, ``A
  fast deep learning method for security vulnerability study of xor pufs,''
  \emph{Electronics}, vol.~9, no.~10, p. 1715, 2020.

\bibitem{wisiolattackers}
N.~Wisiol, C.~Graebnitz, M.~Margraf, M.~Oswald, T.~Soroceanu, and B.~Zengin,
  ``Why attackers lose: Design and security analysis of arbitrarily large xor
  arbiter pufs.''

\bibitem{wisiol2019breaking}
N.~Wisiol, G.~T. Becker, M.~Margraf, T.~A. Soroceanu, J.~Tobisch, and
  B.~Zengin, ``Breaking the lightweight secure puf: Understanding the relation
  of input transformations and machine learning resistance,'' in
  \emph{International Conference on Smart Card Research and Advanced
  Applications}.\hskip 1em plus 0.5em minus 0.4em\relax Springer, 2019, pp.
  40--54.

\bibitem{aseeri2018subspace}
A.~O. Aseeri, Y.~Zhuang, and M.~S. Alkatheiri, ``A subspace pre-learning
  approach to fast high-accuracy machine learning of large xor pufs with
  component-differential challenges,'' in \emph{2018 IEEE International
  Conference on Big Data (Big Data)}.\hskip 1em plus 0.5em minus 0.4em\relax
  IEEE, 2018, pp. 1563--1568.

\bibitem{mursi2021experimental}
K.~T. Mursi and Y.~Zhuang, ``Experimental examination of
  component-differentially-challenged xor puf circuits,'' in \emph{Journal of
  Physics: Conference Series}, vol. 1729, no.~1.\hskip 1em plus 0.5em minus
  0.4em\relax IOP Publishing, 2021, p. 012006.

\bibitem{4261134}
G.~E. Suh and S.~Devadas, ``Physical unclonable functions for device
  authentication and secret key generation,'' in \emph{2007 44th ACM/IEEE
  Design Automation Conference}, 2007, pp. 9--14.

\bibitem{becker2015gap}
G.~T. Becker, ``The gap between promise and reality: On the insecurity of xor
  arbiter pufs,'' in \emph{International Workshop on Cryptographic Hardware and
  Embedded Systems}.\hskip 1em plus 0.5em minus 0.4em\relax Springer, 2015, pp.
  535--555.

\bibitem{majzoobi2008lightweight}
M.~Majzoobi, F.~Koushanfar, and M.~Potkonjak, ``Lightweight secure pufs,'' in
  \emph{2008 IEEE/ACM International Conference on Computer-Aided Design}.\hskip
  1em plus 0.5em minus 0.4em\relax IEEE, 2008, pp. 670--673.

\bibitem{gassend2004identification}
B.~Gassend, D.~Lim, D.~Clarke, M.~Van~Dijk, and S.~Devadas, ``Identification
  and authentication of integrated circuits,'' \emph{Concurrency and
  Computation: Practice and Experience}, vol.~16, no.~11, pp. 1077--1098, 2004.

\bibitem{lim2004extracting}
D.~Lim, ``Extracting secret keys from integrated circuits in master thesis,''
  \emph{Massachusetts Institute of Technology}, 2004.

\bibitem{nguyen2019interpose}
P.~H. Nguyen, D.~P. Sahoo, C.~Jin, K.~Mahmood, U.~R{\"u}hrmair, and M.~van
  Dijk, ``The interpose puf: Secure puf design against state-of-the-art machine
  learning attacks,'' \emph{IACR Transactions on Cryptographic Hardware and
  Embedded Systems}, pp. 243--290, 2019.

\bibitem{tobisch2015scaling}
J.~Tobisch and G.~T. Becker, ``On the scaling of machine learning attacks on
  pufs with application to noise bifurcation,'' in \emph{International Workshop
  on Radio Frequency Identification: Security and Privacy Issues}.\hskip 1em
  plus 0.5em minus 0.4em\relax Springer, 2015, pp. 17--31.

\bibitem{santikellur2019deep}
P.~Santikellur, A.~Bhattacharyay, and R.~S. Chakraborty, ``Deep learning based
  model building attacks on arbiter puf compositions,'' Cryptology ePrint
  Archive, Report 2019/566. 2019. Available online: https~…, Tech. Rep.,
  2019.

\bibitem{mursi2020MPUF_JCM}
K.~T. Mursi and Y.~Zhuang, ``Experimental study of
  component-differentially-challenged xor pufs as security primitives for
  internet-of-things,'' \emph{Journal of Communications}, vol.~15, no.~10,
  2020.

\bibitem{wisiol2020splitting}
N.~Wisiol, C.~M{\"u}hl, N.~Pirnay, P.~H. Nguyen, M.~Margraf, J.-P. Seifert,
  M.~van Dijk, and U.~R{\"u}hrmair, ``Splitting the interpose puf: A novel
  modeling attack strategy,'' \emph{IACR Transactions on Cryptographic Hardware
  and Embedded Systems}, pp. 97--120, 2020.

\bibitem{thapaliya2021machine}
B.~Thapaliya, K.~T. Mursi, and Y.~Zhuang, ``Machine learning-based
  vulnerability study of interpose pufs as security primitives for iot
  networks,'' in \emph{2021 IEEE International Conference on Networking,
  Architecture and Storage (NAS)}.\hskip 1em plus 0.5em minus 0.4em\relax IEEE,
  2021, pp. 1--7.

\bibitem{lim2005extracting}
D.~Lim, J.~W. Lee, B.~Gassend, G.~E. Suh, M.~Van~Dijk, and S.~Devadas,
  ``Extracting secret keys from integrated circuits,'' \emph{IEEE Transactions
  on Very Large Scale Integration (VLSI) Systems}, vol.~13, no.~10, pp.
  1200--1205, 2005.

\bibitem{wisiol2021neural}
N.~Wisiol, K.~T. Mursi, J.-P. Seifert, and Y.~Zhuang, ``Neural-network-based
  modeling attacks on xor arbiter pufs revisited.'' \emph{IACR Cryptol. ePrint
  Arch.}, vol. 2021, p. 555, 2021.

\bibitem{pypuf}
\BIBentryALTinterwordspacing
N.~Wisiol, C.~Gräbnitz, C.~Mühl, B.~Zengin, T.~Soroceanu, N.~Pirnay, and
  K.~T. Mursi, ``{pypuf: Cryptanalysis of Physically Unclonable Functions},''
  2021. [Online]. Available: \url{https://doi.org/10.5281/zenodo.3901410}
\BIBentrySTDinterwordspacing

\bibitem{abadi2016tensorflow}
M.~Abadi, P.~Barham, J.~Chen, Z.~Chen, A.~Davis, J.~Dean, M.~Devin,
  S.~Ghemawat, G.~Irving, M.~Isard \emph{et~al.}, ``Tensorflow: A system for
  large-scale machine learning,'' in \emph{12th $\{$USENIX$\}$ symposium on
  operating systems design and implementation ($\{$OSDI$\}$ 16)}, 2016, pp.
  265--283.

\bibitem{gulli2017deep}
A.~Gulli and S.~Pal, \emph{Deep learning with Keras}.\hskip 1em plus 0.5em
  minus 0.4em\relax Packt Publishing Ltd, 2017.

\bibitem{hori2010quantitative}
Y.~Hori, T.~Yoshida, T.~Katashita, and A.~Satoh, ``Quantitative and statistical
  performance evaluation of arbiter physical unclonable functions on fpgas,''
  in \emph{Reconfigurable Computing and FPGAs (ReConFig), 2010 International
  Conference on}.\hskip 1em plus 0.5em minus 0.4em\relax IEEE, 2010, pp.
  298--303.

\end{thebibliography}

\vspace{12pt}

\end{document}